\documentclass[useAMS,usenatbib]{mn2e}

\usepackage{epsfig}
\usepackage{amsmath}
\usepackage{amssymb}
\usepackage{natbib}
\usepackage{threeparttable} 

\usepackage{epstopdf}

\newcommand{\chan}{\textit{Chandra}}
\newcommand{\swift}{\textit{Swift}}

\newcommand{\xmm}{\textit{XMM-Newton}}
\newcommand{\inte}{\textit{Integral}}

\newcommand{\maxi}{\textit{MAXI}}

\newcommand{\Msun}{\mathrm{M}_{\odot}}
\newcommand{\lum}{\mathrm{erg~s}^{-1}}
\newcommand{\flux}{\mathrm{erg~cm}^{-2}~\mathrm{s}^{-1}}
\newcommand{\cnts}{\mathrm{counts~s}^{-1}}
\newcommand{\mdot}{\mathrm{M_{\odot}~yr}^{-1}}
\newcommand{\nh}{\mathrm{cm^{-2}}}

\newcommand{\source}{CXOGClb J174804.8--244648}
\newcommand{\intename}{IGR J17480--2446}
\newcommand{\exo}{EXO~0748--676}
\newcommand{\ks}{KS~1731--260}
\newcommand{\mxb}{MXB~1659--29}
\newcommand{\xte}{XTE~J1701--462}
\newcommand{\xtejonker}{XTE~J1709--267}

\def \mnras {MNRAS}
\def \apj {ApJ}

\def \apjl {ApJL}
\def \aap {A\&A}

\def \pasj {PASJ}

\title[The heated crust of the Terzan 5 X-ray pulsar]{The accretion-heated crust of the transiently accreting 11~Hz X-ray pulsar in the globular cluster Terzan~5}
\author[N. Degenaar \& R. Wijnands]
{N. Degenaar\thanks{e-mail: degenaar@uva.nl} \& R. Wijnands\\
Astronomical Institute "Anton Pannekoek", 
University of Amsterdam, 
Postbus 94249, 1090 GE Amsterdam, the Netherlands
}

\begin{document}

\date{Accepted 2011 March 25.  Received 2011 March 7.}

\pagerange{\pageref{firstpage}--\pageref{lastpage}} \pubyear{0000}

\maketitle

\label{firstpage}

\begin{abstract} 
We report on a \chan\ Director's Discretionary Time observation of the globular cluster Terzan 5, carried out $\sim 7$~weeks after the cessation of the 2010 outburst of the newly discovered transiently accreting 11 Hz X-ray pulsar. We detect a thermal spectrum that can be fitted with a neutron star atmosphere model with a temperature for an observer at infinity of $kT^{\infty} \sim 100$~eV, and a quiescent thermal bolometric luminosity of $L_q \sim 2\times10^{33}~\lum$ for an assumed distance of 5.5~kpc. The thermal emission is elevated above the quiescent base level measured in 2003 and 2009, i.e., prior to the recent accretion outburst. A likely explanation is that the neutron star crust was significantly heated during the recent accretion episode and needs to cool until it restores thermal equilibrium with the core. Although this has been observed for neutron star low-mass X-ray binaries that undergo accretion episodes of years to decades, it is the first time that evidence for crustal heating is detected for a transient system with a regular outburst duration of weeks. This opens up a new window to study heating and cooling of transiently accreting neutron stars.
\end{abstract}


\section{Introduction}\label{sec:intro}
The dense globular cluster Terzan 5 lies in the bulge of the Milky Way Galaxy at an estimated distance of $D=5.5$~kpc \citep[][]{ortolani2007}. Terzan 5 contains an exceptionally large population of millisecond radio pulsars \citep[e.g.,][]{ransom2005}, and high-spatial resolution \chan\ observations have allowed for the identification of many X-ray point sources that are likely quiescent low-mass X-ray binaries (LMXBs) or cataclysmic variables \citep[][]{heinke2006_terzan5}. 

Neutron stars residing in LMXBs accrete matter from a companion that has a mass $\lesssim1~\Msun$. These are often transient systems for which the X-ray luminosity can brighten orders of magnitude due to a sudden strong increase in the mass-accretion rate onto the compact primary. Such accretion outbursts generally reach 2--10 keV X-ray luminosities of $L_X\sim10^{36-38}~\lum$, and last for several weeks. Afterwards, the system returns to the quiescent state with a typical X-ray luminosity of $L_q\sim10^{31-33}~\lum$, and remains as such for years or decades until it enters a new outburst.

\inte\ bulge scan monitoring observations signalled X-ray activity in the direction of Terzan 5 on 2010 October 10 \citep[][]{bordas2010}. The subsequent detection of type-I X-ray bursts \citep[][]{chenevez2010_terzan5} and coherent 11 Hz pulsations \citep[][]{strohmayer2010} demonstrated that the active source was a neutron star LMXB, whereas \chan\ observations provided an accurate localization \citep{pooley2010}. This new X-ray transient, designated \intename/\source\ (J1748 hereafter), is the second confirmed transient neutron star LMXB in this globular cluster \cite[for details on the other one; see][]{wijnands2002_terzan5,heinke2003}.

Timing of the X-ray pulsations revealed that the neutron star orbits its $\sim 0.4 - 1.5 ~\Msun$ donor in $\sim21$~h \citep[][]{papitto2010}. The neutron star is thought to have a relatively strong magnetic field, as inferred from the detection of a broad iron line in the X-ray spectrum \citep[$B\sim0.7-4.0\times10^{9}$~G;][]{miller2011}, timing of the X-ray pulsations \citep[$B\sim0.2-24\times10^{9}$~G;][]{papitto2010} and burst oscillation studies \citep[$B\gtrsim10^{9}$~G;][]{cavecchi2011}.

\begin{figure}
 \begin{center}
\includegraphics[width=8.0cm]{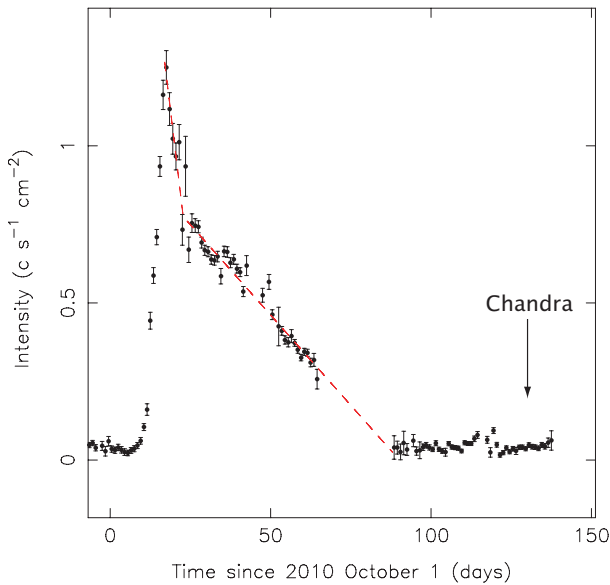}
    \end{center}
\caption[]{{One-day averaged \maxi\ lightcurve of Terzan 5, displaying the 2010 outburst of the recently discovered 11 Hz X-ray pulsar (2--20 keV). The time of our \chan\ observation is indicated. The dashed line represents a fit with a broken linear decay.
}}
 \label{fig:maxi}
\end{figure}

Figure~\ref{fig:maxi} displays the lightcurve of the 2010 outburst of Terzan 5, as observed with the \maxi\ all-sky X-ray monitor mounted on the International Space Station \citep[][]{maxi2009}. The activity commenced around 2010 October 10 \citep[see also][]{bordas2010} and the source soon became very bright, approaching a luminosity of $\sim10^{38}~\lum$ \citep[2--50 keV;][]{altamirano2010_2}. As can be seen in Figure~\ref{fig:maxi}, J1748 remained active until it became unobservable due to Sun angle constraints in 2010 early-December. Terzan 5 was no longer detected when the \maxi\ monitoring resumed in 2010 late-December, implying a 2--30 keV intensity of $\lesssim15$~mCrab \citep[][]{sugizaki2011} or $L_X \lesssim 10^{35}~(D/5.5~\mathrm{kpc})^2~\lum$. This indicates that the accretion activity had likely ceased by that time (see Sect.~\ref{subsec:outburst}).

Analysis of archival \chan\ data revealed that J1748 has a thermal quiescent spectrum \citep[][]{deeg_wijn2011}. Fits to a neutron star atmosphere model yielded a neutron star effective temperature for an observer at infinity of $kT^{\infty} = 72.7 \pm 7.6$~eV, and a quiescent thermal bolometric luminosity of $L_q = (6.2\pm2.5)\times10^{32}~(D/5.5~\mathrm{kpc})^2~\lum$ \citep[][]{deeg_wijn2011}. Such a soft emission component is often seen for neutron star transients in quiescence and is generally interpreted as thermal emission from the stellar surface. The radiated heat is thought to be generated in nuclear reactions occurring deep inside the neutron star crust during accretion outbursts \citep[e.g.,][]{haensel2008}. These maintain the neutron star core at a steady state temperature, thereby producing a stable level of quiescent thermal emission \citep[][]{brown1998}. 

A small sub-group of transient LMXBs undergo accretion outbursts that persist for $>1$~year. Four of such sources were monitored following the cessation of their long X-ray outburst, which revealed that the thermal X-ray emission decayed over the course of years \citep[e.g.,][]{wijnands2001,wijnands2003,cackett2008,cackett2010,degenaar2010_exo2,fridriksson2011,diaztrigo2011}. The observed decrease in thermal radiation has been interpreted as cooling of the neutron star crust, which became severely heated during the prolonged accretion outburst \citep[][]{rutledge2002,wijnands04_quasip}. Confronting the observed cooling curves with neutron star thermal evolution codes gives information about the amount of heat release and thermal conductivity of the neutron star crust, as well as the properties of the stellar core \citep[][]{shternin07,brown08}. 

The unusually long outburst duration of these quasi-persistent neutron star LMXBs provides the necessary conditions to significantly lift the neutron star crust temperature so that the thermal relaxation becomes observable. However, \citet{brown1998} argued that in regular transients with outburst durations of weeks, the crust might also become significantly heated above the core temperature, provided that the outburst is bright enough compared to the quiescent base level. As noted by \citet[][]{deeg_wijn2011}, the high outburst luminosity and relatively faint quiescent level of J1748 would then make it a good candidate to search for neutron star crust cooling.

In this letter, we report on a \chan\ Director's Discretionary Time (DDT) observation of the globular cluster Terzan 5, obtained after the cessation of the 2010 outburst of the newly discovered 11 Hz X-ray pulsar. The aim of this observation was to study the effect of the bright accretion outburst on the thermal properties of the neutron star crust.

\section{Observations, analysis and results}\label{sec:results}

\begin{figure*}
 \begin{center}
\includegraphics[width=8.0cm]{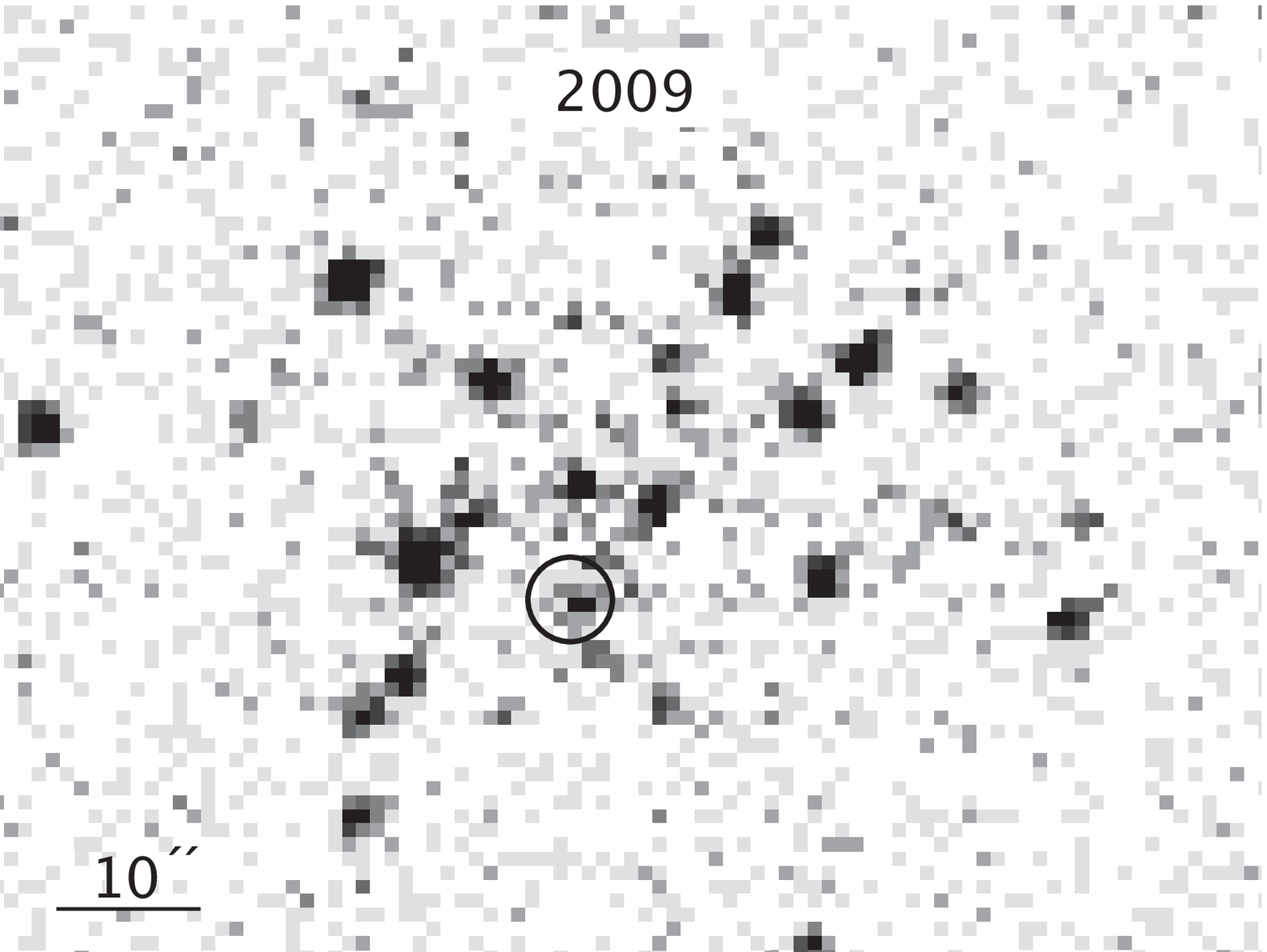}\hspace{0.5cm}
\includegraphics[width=8.0cm]{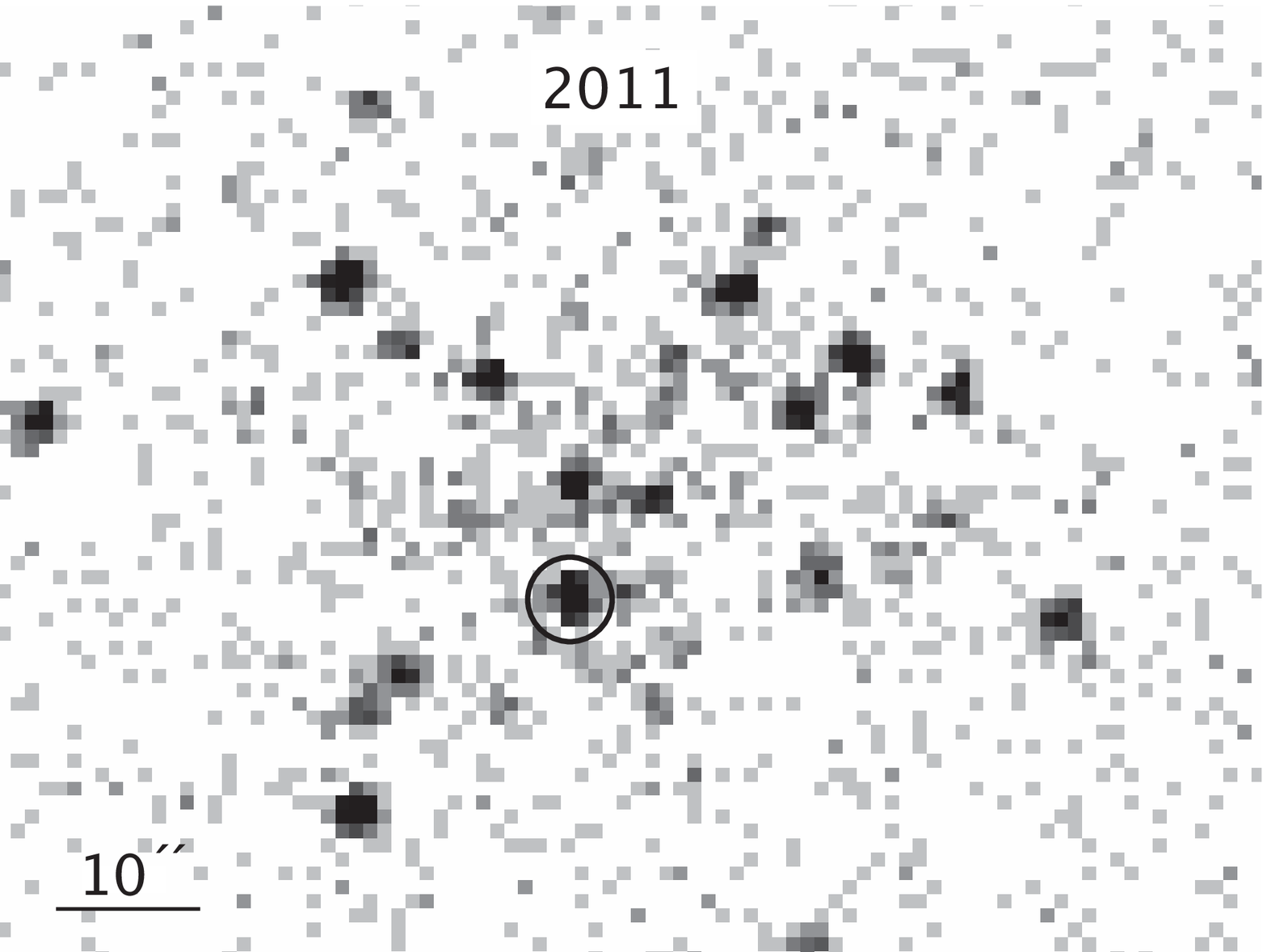}
    \end{center}
\caption[]{{\chan/ACIS images of the Terzan 5 cluster core, indicating the 11 Hz X-ray pulsar that went into outburst in 2010 (circle). Left: Pre-outburst image obtained on 2009 July 15--16, $\sim64$~weeks prior to the 2010 accretion activity. Right: Observation performed $\sim 7$ weeks post-outburst, on 2011 February 17.
}}
 \label{fig:ds9}
\end{figure*}

Terzan 5 was observed with the \chan/ACIS-S on 2011 February 17 from 09:06--17:58 \textsc{ut} for 29.7 ks (ID 13225). The cluster is positioned on the S3 chip and the data was obtained in the faint mode with the nominal frame time of 3.2~s. Figure~\ref{fig:ds9} displays an image of the 2011 \chan\ data, together with a $\sim36.4$~ks archival observation that was obtained on 2009 July 15--16 \citep[see][]{deeg_wijn2011}. It is clear from these images that J1748 is brighter in 2011, $\sim2$~months after the cessation of the 2010 activity, than it was in 2009, $\sim 14$ months prior to that outburst. 

For the purpose of directly comparing the new \chan\ data with archival observations, we use the same analysis and reduction steps as outlined in \citet{deeg_wijn2011}, employing the \textsc{ciao} software tools (v. 4.2). No background flares occurred during the observation, so all data was used in further analysis. Source count rates and lightcurves were extracted from a $1''$ circular region, centred at the source position, using the tool \textsc{dmextract}. Corresponding background events were obtained from a circular region with a radius of $40''$, positioned on a source-free part of the CCD that was located $\sim1.4'$ west of the cluster core. 

Our target is detected at a count rate of $(6.5\pm 0.5)\times10^{-3}~\cnts$, which is about 6 times higher than observed in archival data obtained in 2003 and 2009 \citep[see][]{deeg_wijn2011}. A total of 192 net source photons were collected. We obtained source and background spectra using the tool \textsc{psextract}, and generated redistribution matrices (rmf) and ancillary response files (arf) with the tasks \textsc{mkacisrmf} and \textsc{mkarf}, respectively. We group the spectrum to contain a minimum of 20 photons per bin, and fit the data in the 0.5--8 keV energy range with \textsc{Xspec} (v. 12.6). 

\begin{figure}
 \begin{center}
\includegraphics[width=8.0cm]{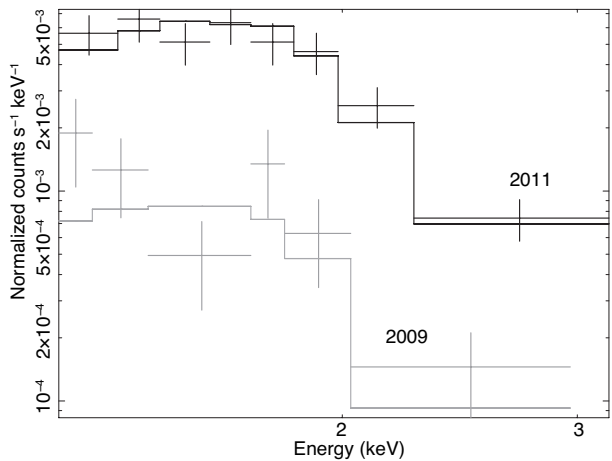}
    \end{center}
\caption[]{{\chan/ACIS spectra of the 11 Hz X-ray pulsar in Terzan 5, as obtained in 2009 (grey) and 2011 (black). The solid lines represent fits to a neutron star atmosphere model.}}
 \label{fig:spec}
\end{figure}

Figure~\ref{fig:spec} displays the spectrum of J1748, as obtained from our 2011 \chan\ observation. For comparison, the 2009 spectral data is also shown. Both spectra are soft, with most photons detected below $\sim2$~keV. The 2011 spectral data is fitted with an absorbed neutron star atmosphere model \textsc{nsatmos} \citep{heinke2006}. We keep the distance model parameter at $D=5.5$~kpc \citep[][]{ortolani2007}, whereas the neutron star mass and radius are fixed at $M_{\mathrm{NS}}=1.4~\Msun$ and $R_{\mathrm{NS}}=10$~km. The model normalization, reflecting the fraction of the surface that is emitting, is set to the recommended value of 1. The only free fit parameters are the hydrogen column density ($N_H$) and the neutron star effective temperature. All errors reported in this work refer to $90\%$ confidence intervals and the interstellar hydrogen absorption is accounted for by employing the \textsc{phabs} model.

Fitting the 2011 spectral data yields $N_H=(2.0\pm 0.4)\times10^{22}~\nh$ and $kT^{\infty}=98.2\pm 4.5$~eV, with a reduced chi-square value of $\chi^2_{\nu}=0.63$ for 6 degrees of freedom (dof). The corresponding 0.5--10 keV unabsorbed flux is $(4.7\pm 1.0)\times10^{-13}~\flux$, and the associated luminosity is $(1.7\pm 0.4)\times10^{33}~(D/5.5~\mathrm{kpc})^2~\lum$. Extrapolation of the model fit to the 0.01--100 keV energy range yields an estimate of the thermal bolometric flux of $(6.9\pm 1.2)\times10^{-13}~\flux$. This implies a thermal bolometric luminosity of $L_q = (2.5\pm 0.5)\times10^{33}~(D/5.5~\mathrm{kpc})^2~\lum$. The \textsc{nsatmos} model fit is plotted in Figure~\ref{fig:spec}, which demonstrates that the spectral data is described well by thermal emission alone. Adding a powerlaw with index $\Gamma=1-2$, as is often detected for neutron star transients in quiescence, does not improve the fit. This leads us to conclude that any hard spectral component contributes less than $\sim15\%$ to the total unabsorbed 0.5--10 keV flux.

The hydrogen column density obtained when fitting the 2011 \chan\ observation is slightly lower than the value retrieved from fits to the 2003 and 2009 spectral data \citep[$N_H=2.1\pm0.9 \times10^{22}~\nh$;][]{deeg_wijn2011}, although consistent within the uncertainties. To measure the difference in temperature and thermal flux, we re-fitted all data simultaneously, with $N_H$ tied between the different observations. The 2003 and 2009 data were treated as a single spectrum to improve statistics, which is justified because no spectral differences were found between the two observations \citep[][]{deeg_wijn2011}. For the simultaneous fit we find $N_H=(2.0\pm0.3) \times10^{22}~\nh$, while $kT^{\infty}=72.1\pm3.9$ and $100.8\pm4.3$~eV for the 2003/2009 and 2011 data, respectively ($\chi^2_{\nu}=0.75$ for 16 dof). The corresponding 0.5--10 keV unabsorbed luminosities are $(3.3\pm0.9)\times10^{32}$ and $(16.0\pm3.1)\times10^{32}~(D/5.5~\mathrm{kpc})^2~\lum$, while the inferred thermal bolometric luminosities are $(5.9\pm1.4)\times10^{32}$ and $(22.8\pm3.9)\times10^{32}~(D/5.5~\mathrm{kpc})^2~\lum$. 

We note that the employed \textsc{nsatmos} model does not take into account any effects of the magnetic field on the emerging photospheric radiation \citep[][]{heinke2006}. This is justified for sources with estimated magnetic fields of $B\lesssim10^{9}$~G, but may not be a valid assumption for the 11 Hz pulsar in Terzan 5 (see Section~\ref{sec:intro}). However, magnetized neutron star atmosphere models incorporate field strengths that are well in excess of the estimates for J1748 ($B\gtrsim10^{12}$~G, e.g., \textsc{nsa}, \textsc{nsamax}). Nevertheless, we make a direct comparison between the observations by using the same spectral model, so the fractional change in neutron star temperature and bolometric luminosity as inferred in this work is robust and not caused by any model uncertainties.

\section{Discussion}\label{sec:discuss}
We report on the spectral properties of the newly discovered 11 Hz X-ray pulsar \intename/\source\ in the globular cluster Terzan 5, as observed with \chan\ within $\sim 7$ weeks (see below) after the cessation of its 2010 accretion outburst. The quiescent spectrum is dominated by thermal emission that fits to a neutron star atmosphere model with $kT^{\infty} \sim 100$~eV for an inferred thermal bolometric luminosity of $L_q \sim 2 \times 10^{33}~(D/5.5~\mathrm{kpc})^2~\lum$. Archival \chan\ observations performed in 2003 and 2009 reveal quiescent spectra that fit to a neutron star atmosphere model with $kT^{\infty} \sim 72$~eV, yielding a thermal bolometric luminosity of $L_q \sim 6 \times 10^{32}~(D/5.5~\mathrm{kpc})^2~\lum$ \citep{deeg_wijn2011}. 

Our new \chan\ observation demonstrates that within two months after the cessation of the recent accretion outburst, the thermal flux and neutron star temperature are elevated above the quiescent base level measured in 2003 and 2009. In analogy with that seen for quasi-persistent LMXBs, we attribute this to heating of the neutron star crust due to its bright 2010 accretion outburst. If true, this leads to the clear prediction that the crust is expected to cool down in the next months, until thermal equilibrium with the core is re-established and the neutron star returns to its quiescent base level. \chan\ observations carried out within the next year are thus expected to detect a decrease in neutron star effective temperature and thermal luminosity, down to the values measured in archival data.

An alternative explanation that may be invoked is that the elevated emission level is caused by residual accretion. It has been proposed that low-level accretion onto the neutron star surface produces thermal radiation \citep[][]{zampieri1995}, and the X-ray spectrum may appear indistinguishable from a passive neutron star atmosphere \citep[][]{soria2011}. However, there are also indications that low-level accretion involves a hard spectral component. One example is \xte, which has been extensively monitored in quiescence with \chan, \xmm\ and \swift\ since its 1.5-yr long outburst ended in 2007 \citep[][]{fridriksson2010,fridriksson2011}. This source exhibits X-ray flares, suggestive of (sporadic) low-level accretion, which are associated with a strong increase in the powerlaw spectral component \citep[][]{fridriksson2011}. In case of J1748, we detect a purely thermal spectrum, both in the new \chan\ observation and in archival data. Any possible hard spectral component contributes $\lesssim15\%$ to the total unabsorbed 0.5--10 keV flux. 

Furthermore, the observed quiescent thermal emission may vary a factor of $\sim2-3$ from one quiescent epoch to another, due to the changing amount of fuel and hydrogen/helium abundances present after each outburst. This influences the fraction of heat conducted from the crust towards the core and surface, and hence the observed thermal quiescent emission \citep[][]{brown2002}. If this is the case for J1748, the quiescent thermal luminosity is expected to remain at the value inferred from our 2011 \chan\ observation, provided that the accretion has switched off completely. 

We consider crustal heating the most likely explanation for the observed elevated quiescent emission of the 11 Hz X-ray pulsar in Terzan 5, but additional \chan\ observations are required to rule out the above alternative scenarios.

\subsection{Outburst constraints}\label{subsec:outburst}
For the purpose of calculating crust cooling curves, it is necessary to constrain the end of the accretion outburst. We therefore try to determine the time at which the source intensity dropped below the detection threshold of \maxi\ (implying a 2--30 keV luminosity of $L_X \sim10^{35}~\lum$). Unfortunately, Sun angle constraints deprived our view of Terzan 5 in 2010 December, so that the end of the outburst was not observed (see Fig.~\ref{fig:maxi}). The outburst commenced around 2010 October 10 and the source was still active on 2010 December 4. This indicates a minimum outburst duration of 7 weeks (55 days). Since no activity was detected when the \maxi\ monitoring observations resumed on 2010 December 28, the outburst duration is constrained to be $<11$ weeks (79 days). Our \chan\ observation carried out on 2011 February 17, thus took place between 7--10 weeks (51--75 days) after the cessation of the accretion outburst.

We try to refine the estimate of the endpoint of the outburst by fitting the \maxi\ lightcurve to a decay function. After several trials, we found that a broken linear function provides the best description of the decaying part of the lightcurve (see Fig.~\ref{fig:maxi}). By extrapolating this fit to later times, we tentatively place the end of the outburst around 2010 December 26 (MJD 55556), although this estimate is subject to the uncertainty in the shape of the outburst decay (e.g., the decay may have accelerated at some point). This suggests that the outburst had a duration of $\sim11$~weeks ($\sim77$~days) and that our \chan\ observation was thus performed $\sim7$ weeks ($\sim53$~days) after the cessation of the 2010 outburst. The estimated average mass-accretion rate during outburst is $\dot{M}\sim3\times10^{-9}~\mdot$ \citep[][]{cavecchi2011,deeg_wijn2011}.

\subsection{Comparison with other sources}
We have observed J1748 approximately $7$~weeks after the cessation of its 2010 accretion outburst, and found that the thermal bolometric luminosity is elevated by a factor $\sim4$ above the quiescent base level, while the inferred neutron star effective temperature is higher by a factor of $\sim1.4$. This is in between the values obtained for the four quasi-persistent neutron star LMXBs that were observed on similar timescales following the cessation of their most recent outbursts: \ks\ \citep[][]{cackett2010}, \mxb\ \citep[][]{cackett2008}, \xte\ \citep[][]{fridriksson2011} and \exo\ \citep[][]{degenaar2010_exo2,diaztrigo2011}. Our results on J1748 thus seems to fit in with the crust cooling detected from the quasi-persistent LMXBs.

Amongst the transient systems with regular outburst durations, Aql X-1 was observed with \chan\ four times after the end of its 2000 activity. Although the initial two data points suggested a $\sim50\%$ flux decay, the source intensity had increased in the subsequent observations \citep[][]{rutledge2002_aqlX1}. Since the quiescent source flux has been found to vary up to a factor of a few \citep[][]{cackett2011}, there is no strong case to ascribe the observed initial decay in quiescent flux to cooling of the accretion-heated crust. Moreover, \citet{brown1998} argued that the relatively high quiescent luminosity, compared to the outburst level, would raise the neutron star crust temperature in Aql X-1 by $\lesssim1\%$, rendering it unlikely that cooling of the neutron star crust can actually be observed for this particular source. 

Another source, \xtejonker, was monitored with \chan\ several times during the decay of its bright 2002 outburst \citep[][]{jonker2003}. A follow-up pointing obtained $\sim1$~year later detected the source at a 0.5--10 keV intensity that was a factor $\sim2-3$ lower than found $\sim1$~month after the outburst \citep[][]{jonker2004}. These authors noted that the observed flux decay after the outburst was much lower than found for \mxb, which decreased by a factor of $\sim7-9$ in a similar time span \citep[][]{wijnands2004}. However, the observations of \xtejonker\ are comparable to our results for J1748, so possibly the neutron star crust temperature was elevated after the bright accretion outburst for this source as well. 

\subsection{Concluding remarks}
The results presented in this work suggest that it is possible to detect the effects of crustal heating from transient neutron star LMXBs having regular outburst durations of weeks. This opens up an additional sample of potential targets that can be used for such studies. Promising sources to search for crust cooling once a new accretion episode has passed are those that become bright during outburst, but have a relatively low quiescent luminosity, such as e.g., Cen X-4 \citep[][]{rutledge1999,campana2004} and 2S 1803--245 \citep[][]{cornelisse2007}. It will be interesting to probe the differences in crust cooling curves of regular transient LMXBs and their quasi-persistent relatives. Another exciting implication is the possibility to compare the crust cooling curves of a single source after outbursts with different lengths and peak luminosities. In quasi-persistent LMXBs the neutron star crust is expected to be close to a thermal steady-state profile during the long outburst \citep[][]{brown08}, whereas in transients with regular outburst durations the thermal profile in the crust, and hence the resulting cooling curve, can differ from one outburst to another. This offers new ways to study heating and cooling processes in transiently accreting neutron stars. \\

\noindent 
{\bf Acknowledgements.}\\
The authors are grateful to Harvey Tananbaum and the \chan\ science team for making this DDT observation possible. This research made use of the \maxi\ data provided by RIKEN, JAXA and the MAXI team. Our work was supported by the Netherlands Research School for Astronomy (NOVA). RW acknowledges support from a European Research Council (ERC) starting grant.\\

\bibliographystyle{mn2e}

\begin{thebibliography}{}

\bibitem[\protect\citeauthoryear{{Altamirano et al.}}{{Altamirano et
  al.}}{2010}]{altamirano2010_2}
{Altamirano et al.} D.,  2010, The Astronomer's Telegram, 2946

\bibitem[\protect\citeauthoryear{{Bordas et al.}}{{Bordas et
  al.}}{2010}]{bordas2010}
{Bordas et al.} P.,  2010, The Astronomer's Telegram, 2919

\bibitem[\protect\citeauthoryear{{Brown}, {Bildsten} \& {Chang}}{{Brown}
  et~al.}{2002}]{brown2002}
{Brown} E.~F.,  {Bildsten} L.,    {Chang} P.,  2002, \apj, 574, 920

\bibitem[\protect\citeauthoryear{{Brown}, {Bildsten} \& {Rutledge}}{{Brown}
  et~al.}{1998}]{brown1998}
{Brown} E.~F.,  {Bildsten} L.,    {Rutledge} R.~E.,  1998, ApJ Lett, 504, L95

\bibitem[\protect\citeauthoryear{{Brown} \& {Cumming}}{{Brown} \&
  {Cumming}}{2009}]{brown08}
{Brown} E.~F.,  {Cumming} A.,  2009, \apj, 698, 1020

\bibitem[\protect\citeauthoryear{{Cackett}, {Brown}, {Cumming}, {Degenaar},
  {Miller} \& {Wijnands}}{{Cackett} et~al.}{2010}]{cackett2010}
{Cackett} E.~M.,  {Brown} E.~F.,  {Cumming} A.,  {Degenaar} N.,  {Miller}
  J.~M.,    {Wijnands} R.,  2010, \apjl, 722, L137

\bibitem[\protect\citeauthoryear{{Cackett}, {Fridriksson}, {Homan}, {Miller} \&
  {Wijnands}}{{Cackett} et~al.}{2011}]{cackett2011}
{Cackett} E.~M.,  {Fridriksson} J.~K.,  {Homan} J.,  {Miller} J.~M.,
  {Wijnands} R.,  2011, \mnras\ in press, arXiv:1102.5016

\bibitem[\protect\citeauthoryear{{Cackett}, {Wijnands}, {Miller}, {Brown} \&
  {Degenaar}}{{Cackett} et~al.}{2008}]{cackett2008}
{Cackett} E.~M.,  {Wijnands} R.,  {Miller} J.~M.,  {Brown} E.~F.,    {Degenaar}
  N.,  2008, \apjl, 687, L87

\bibitem[\protect\citeauthoryear{{Campana}, {Israel}, {Stella}, {Gastaldello} \&
  {Mereghetti}}{{Campana} et~al.}{2004}]{campana2004}
{Campana}, S., {Israel}, G.~L., {Stella}, L., {Gastaldello}, F., {Mereghetti}, S., 2004, \apj, 601, 474

\bibitem[\protect\citeauthoryear{{Cavecchi} {et al.}}{{Cavecchi}
  et~al.}{2011}]{cavecchi2011}
{Cavecchi} Y.,  et al.,  2011, \apjl\ submitted, arXiv:1102.1548

\bibitem[\protect\citeauthoryear{{Chenevez et al.}}{{Chenevez et
  al.}}{2010}]{chenevez2010_terzan5}
{Chenevez et al.} J.,  2010, The Astronomer's Telegram, 2924

\bibitem[\protect\citeauthoryear{{Cornelisse}, {Wijnands} \&
  {Homan}}{{Cornelisse} et~al.}{2007}]{cornelisse2007}
{Cornelisse} R.,  {Wijnands} R.,    {Homan} J.,  2007, \mnras, 380, 1637

\bibitem[\protect\citeauthoryear{{Degenaar} \& {Wijnands}}{{Degenaar} \&
  {Wijnands}}{2011}]{deeg_wijn2011}
{Degenaar} N.,  {Wijnands} R.,  2011, \mnras, 412, L68

\bibitem[\protect\citeauthoryear{{Degenaar} {et al.}}{{Degenaar}
  et~al.}{2010}]{degenaar2010_exo2}
{Degenaar} N., et al., 2011, \mnras, 412, 1409

\bibitem[\protect\citeauthoryear{{Diaz Trigo}, {Boirin}, {Costantini}, {M{\'e}ndez}
  \& {Parmar}}{{Diaz Trigo} et~al.}{2011}]{diaztrigo2011}
{Diaz Trigo} M.,  {Boirin} L.,  {Costantini} E.,  {Mendez} M., {Parmar} A.,
  2011, \aap, 528, 150

\bibitem[\protect\citeauthoryear{{Fridriksson} {et al.}}{{Fridriksson}
  et~al.}{2011}]{fridriksson2011}
{Fridriksson} J.~K.,  et al., 2011, \apj\ submitted, arXiv:1101.0081

\bibitem[\protect\citeauthoryear{{Fridriksson} {et al.}}{{Fridriksson} et~al.}{2010}]{fridriksson2010}
{Fridriksson} J.~K.,  et al., 2010, \apj, 714, 270

\bibitem[\protect\citeauthoryear{{Haensel} \& {Zdunik}}{{Haensel} \&
  {Zdunik}}{2008}]{haensel2008}
{Haensel} P.,  {Zdunik} J.~L.,  2008, \aap, 480, 459

\bibitem[\protect\citeauthoryear{{Heinke}, {Edmonds}, {Grindlay}, {Lloyd},
  {Cohn} \& {Lugger}}{{Heinke} et~al.}{2003}]{heinke2003}
{Heinke} C.~O.,  {Edmonds} P.~D.,  {Grindlay} J.~E.,  {Lloyd} D.~A.,  {Cohn}
  H.~N.,    {Lugger} P.~M.,  2003, \apj, 590, 809

\bibitem[\protect\citeauthoryear{{Heinke}, {Rybicki}, {Narayan} \&
  {Grindlay}}{{Heinke} et~al.}{2006}]{heinke2006}
{Heinke} C.~O.,  {Rybicki} G.~B.,  {Narayan} R.,    {Grindlay} J.~E.,  2006,
  \apj, 644, 1090

\bibitem[\protect\citeauthoryear{{Heinke}, {Wijnands}, {Cohn}, {Lugger},
  {Grindlay}, {Pooley} \& {Lewin}}{{Heinke} et~al.}{2006}]{heinke2006_terzan5}
{Heinke} C.~O.,  {Wijnands} R.,  {Cohn} H.~N.,  {Lugger} P.~M.,  {Grindlay}
  J.~E.,  {Pooley} D.,    {Lewin} W.~H.~G.,  2006, \apj, 651, 1098

\bibitem[\protect\citeauthoryear{{Jonker}, {Galloway}, {McClintock}, {Buxton},
  {Garcia} \& {Murray}}{{Jonker} et~al.}{2004}]{jonker2004}
{Jonker} P.~G.,  {Galloway} D.~K.,  {McClintock} J.~E.,  {Buxton} M.,  {Garcia}
  M.,    {Murray} S.,  2004, \mnras, 354, 666

\bibitem[\protect\citeauthoryear{{Jonker}, {M{\'e}ndez}, {Nelemans}, {Wijnands}
  \& {van der Klis}}{{Jonker} et~al.}{2003}]{jonker2003}
{Jonker} P.~G.,  {M{\'e}ndez} M.,  {Nelemans} G.,  {Wijnands} R.,    {van der
  Klis} M.,  2003, \mnras, 341, 823

\bibitem[\protect\citeauthoryear{{Matsuoka et al.}}{{Matsuoka et
  al.}}{2009}]{maxi2009}
{Matsuoka et al.} M.,  2009, \pasj, 61, 999

\bibitem[\protect\citeauthoryear{{Miller}, {Maitra}, {Cackett}, {Bhattacharyya}
  \& {Strohmayer}}{{Miller} et~al.}{2011}]{miller2011}
{Miller} J.~M.,  {Maitra} D.,  {Cackett} E.~M.,  {Bhattacharyya} S.,
  {Strohmayer} T.~E.,  2011, \apjl, 731, L7

\bibitem[\protect\citeauthoryear{{Ortolani}, {Barbuy}, {Bica}, {Zoccali} \&
  {Renzini}}{{Ortolani} et~al.}{2007}]{ortolani2007}
{Ortolani} S.,  {Barbuy} B.,  {Bica} E.,  {Zoccali} M.,    {Renzini} A.,  2007,
  \aap, 470, 1043

\bibitem[\protect\citeauthoryear{{Papitto}, {D'A{\`i}}, {Motta}, {Riggio},
  {Burderi}, {di Salvo}, {Belloni} \& {Iaria}}{{Papitto}
  et~al.}{2011}]{papitto2010}
{Papitto} A.,  {D'A{\`i}} A.,  {Motta} S.,  {Riggio} A.,  {Burderi} L.,  {di
  Salvo} T.,  {Belloni} T.,    {Iaria} R.,  2011, \aap, 526, L3

\bibitem[\protect\citeauthoryear{{Pooley}, {Homan}, {Heinke}, {Linares},
  {Altamirano} \& {Lewin}}{{Pooley} et~al.}{2010}]{pooley2010}
{Pooley} D.,  {Homan} J.,  {Heinke} C.,  {Linares} M.,  {Altamirano} D.,
  {Lewin} W.,  2010, The Astronomer's Telegram, 2974

\bibitem[\protect\citeauthoryear{{Ransom}, {Hessels}, {Stairs}, {Freire},
  {Camilo}, {Kaspi} \& {Kaplan}}{{Ransom} et~al.}{2005}]{ransom2005}
{Ransom} S.~M.,  {Hessels} J.~W.~T.,  {Stairs} I.~H.,  {Freire} P.~C.~C.,
  {Camilo} F.,  {Kaspi} V.~M.,    {Kaplan} D.~L.,  2005, Science, 307, 892

\bibitem[\protect\citeauthoryear{{Rutledge}, {Bildsten}, {Brown}, {Pavlov} \&
  {Zavlin}}{{Rutledge} et~al.}{1999}]{rutledge1999}
{Rutledge} R.~E.,  {Bildsten} L.,  {Brown} E.~F.,  {Pavlov} G.~G.,    {Zavlin}
  V.~E.,  1999, \apj, 514, 945

\bibitem[\protect\citeauthoryear{{Rutledge}, {Bildsten}, {Brown}, {Pavlov} \&
  {Zavlin}}{{Rutledge} et~al.}{2002}]{rutledge2002_aqlX1}
{Rutledge} R.~E.,  {Bildsten} L.,  {Brown} E.~F.,  {Pavlov} G.~G.,    {Zavlin}
  V.~E.,  2002, \apj, 577, 346

\bibitem[\protect\citeauthoryear{{Rutledge}, {Bildsten}, {Brown}, {Pavlov},
  {Zavlin} \& {Ushomirsky}}{{Rutledge} et~al.}{2002}]{rutledge2002}
{Rutledge} R.~E.,  {Bildsten} L.,  {Brown} E.~F.,  {Pavlov} G.~G.,  {Zavlin}
  V.~E.,    {Ushomirsky} G.,  2002, ApJ, 580, 413

\bibitem[\protect\citeauthoryear{{Shternin}, {Yakovlev}, {Haensel} \&
  {Potekhin}}{{Shternin} et~al.}{2007}]{shternin07}
{Shternin} P.~S.,  {Yakovlev} D.~G.,  {Haensel} P.,    {Potekhin} A.~Y.,  2007,
  \mnras, 382, L43

\bibitem[\protect\citeauthoryear{{Soria}, {Zampieri}, {Zane} \& {Wu}}{{Soria}
  et~al.}{2011}]{soria2011}
{Soria} R.,  {Zampieri} L.,  {Zane} S.,    {Wu} K.,  2011, \mnras, 410, 1886

\bibitem[\protect\citeauthoryear{{Strohmayer}, {Markwardt}, {Pereira} \&
  {Smith}}{{Strohmayer} et~al.}{2010}]{strohmayer2010}
{Strohmayer} T.~E.,  {Markwardt} C.~B.,  {Pereira} D.,    {Smith} E.~A.,  2010,
  The Astronomer's Telegram, 2929

\bibitem[\protect\citeauthoryear{{Sugizaki et al.}}{{Sugizaki et
  al.}}{2011}]{sugizaki2011}
{Sugizaki et al.} M.,  2011, \pasj\ in press, arXiv:1102.0891

\bibitem[\protect\citeauthoryear{{Wijnands}}{{Wijnands}}{2004}]{wijnands04_quasip}
{Wijnands} R.,  2004, ArXiv e-prints astro-ph/0405089

\bibitem[\protect\citeauthoryear{{Wijnands}, {Homan} \& {Remillard}}{{Wijnands}
  et~al.}{2002}]{wijnands2002_terzan5}
{Wijnands} R.,  {Homan} J., {Remillard} R.,  2002, The Astronomer's
  Telegram, 101 

\bibitem[\protect\citeauthoryear{{Wijnands}, {Homan},{Miller} \& {Lewin}}{{Wijnands}
  et~al.}{2004}]{wijnands2004}
{Wijnands}, R., {Homan}, J., {Miller}, J.~M., {Lewin}, W.~H.~G,  2004, \apjl, 606, L61

\bibitem[\protect\citeauthoryear{{Wijnands}, {Miller}, {Markwardt}, {Lewin} \&
  {van der Klis}}{{Wijnands} et~al.}{2001}]{wijnands2001}
{Wijnands} R.,  {Miller} J.~M.,  {Markwardt} C.,  {Lewin} W.~H.~G.,    {van der
  Klis} M.,  2001, ApJ Lett, 560, L159

\bibitem[\protect\citeauthoryear{{Wijnands}, {Nowak}, {Miller}, {Homan},
  {Wachter} \& {Lewin}}{{Wijnands} et~al.}{2003}]{wijnands2003}
{Wijnands} R.,  {Nowak} M.,  {Miller} J.~M.,  {Homan} J.,  {Wachter} S.,
  {Lewin} W.~H.~G.,  2003, ApJ, 594, 952

\bibitem[\protect\citeauthoryear{{Zampieri}, {Turolla}, {Zane} \&
  {Treves}}{{Zampieri} et~al.}{1995}]{zampieri1995}
{Zampieri} L.,  {Turolla} R.,  {Zane} S.,    {Treves} A.,  1995, \apj, 439, 849

\end{thebibliography}

\label{lastpage}
\end{document}